\documentclass[aps,pre,amsmath,lengthcheck,superscriptaddress,onecolumn]{revtex4-2}

\usepackage{graphicx}\graphicspath{ {figures/} }
\usepackage{hyperref}
\hypersetup{colorlinks,allcolors=blue,breaklinks}

\newcommand{\be}{\begin{equation}}
\newcommand{\ee}{\end{equation}}
\newcommand{\ba}{\begin{align}}
\newcommand{\ea}{\end{align}}
\newcommand{\bi}{\begin{itemize}}
\newcommand{\ei}{\end{itemize}}

\newcommand{\bla}{bla\\bla\\bla\\bla\\bla}

\begin{document}

\title{Nonlinear response of the irreversible work via generalized relaxation functions}

\author{Pierre Naz\'e}
\email{pierre.naze@unesp.br}

\affiliation{\it Universidade Estadual Paulista, 14800-090, Araraquara, S\~ao Paulo, Brazil}

\date{\today}

\begin{abstract}

The nonlinear response of the excess work, when made via series expansion in the parameter perturbation of the average thermodynamic work, requires adjustments to agree with the Second Law of Thermodynamics. In this work, I present a well-behaved nonlinear response of the irreversible work, based on its well-known cumulant series expansion. From the generalization of the fluctuation-dissipation relation derived from it, I define the terms of the series expansion in the parameter perturbation of the irreversible work by the terms of the cumulants. Since every cumulant depends on raw moments, I define from them the generalized relaxation functions, whose arbitrary constants were chosen guaranteeing the accomplishment of the Second Law of Thermodynamics. A procedure to calculate the nonlinear response of the irreversible work is then provided.  

\end{abstract}

\maketitle

\section{Introduction}

The theoretical calculation of the irreversible work is a hard task to accomplish since the probabilistic distribution of the phase space is not completely known for arbitrary driven processes. A partial solution for this problem is considering approximations, like linear response theory, where the parameter perturbation is slightly changed. In this case, the only requirement is knowing the solution of the non-perturbed system. Higher orders of such theory undoubtedly would be desired to have a better range of performance in the parameter perturbation.

In Ref.~\cite{naze2022series} the development of the nonlinear response theory was made considering thermally isolated systems. Unfortunately, the final result was disappointing since an {\it ad hoc} procedure was necessary for the series to agree with the Second Law of Thermodynamics. Indeed, without such intervention, the excess work would diverge for large switching times. This is a sign that this nonlinear response theory captures this usual behavior that occurs in time-independent perturbation theory in classical mechanics.  

In this work, I present a nonlinear response theory for systems performing isothermal processes. In order to avoid any possible divergence for long switching times, I adopt an alternative procedure to the calculation of the irreversible work: the well-known expansion in cumulants~\cite{jarzynski1997}. In this manner, the irreversible work can be expressed in terms involving raw moments of the distribution, which are written in their turn in generalized relaxation functions. To agree with the Second Law of Thermodynamics, the arbitrary constants on which the generalized relaxation functions depend are chosen for the moments to nullify for large switching times. This immediately implies that the cumulants will have the same behavior, and so will the terms of the irreversible work. This will avoid the diverging behavior observed for thermally isolated systems. A procedure for the calculation of the irreversible work using this nonlinear response theory is provided at the end.

\section{Preliminaries}

Consider a classical system weakly coupled with a heat bath of temperature $\beta^{-1}$. Their Hamiltonians are $\mathcal{H}_{\mathcal{S}}$ and $\mathcal{H}_{\mathcal{B}}$, such that their sum are represented by $\mathcal{H}_0$. A linear perturbation $\lambda(t)\mathcal{H}_1$ is added to the composed system, where $\lambda(t)=\lambda_0+g(t/\tau)\delta\lambda$. Here $\tau$ is the switching time of the process, $\delta\lambda$ the parameter perturbation, and $g(0)=0$ and $g(\tau)=1$. Our interest is to calculate the irreversible work
\be
W_{\rm irr}(\tau)=\langle W\rangle-\Delta F,
\ee
where $\langle W\rangle$ is the average thermodynamic work
\be
\langle W\rangle=\int_0^\tau \langle \mathcal{H}_1\rangle(t)\dot{\lambda}(t)dt,
\ee
and $\Delta F$ is the difference of Helmholtz's free energy between the final and initial equilibrium state of the composed system. The average is calculated in the probability distribution $\rho$, which satisfies Liouville's equation
\be
\frac{\partial\rho}{\partial t} = \mathcal{L}\rho,
\ee
where $\mathcal{L}(\bullet)=-\{\bullet,\mathcal{H}_0+\lambda(t)\mathcal{H}_1\}$ is called Liouville's operator.

\section{Cumulant series expansion}

It is well known that the irreversible work can be defined by the series expansion~\cite{jarzynski1997}
\be
\beta W_{\rm irr} = \sum_{n=2}^{\infty}\beta\omega_n,
\ee
with
\be
\beta\omega_n = \frac{(-\beta)^n}{n!}\kappa_n,
\label{eq:omegan}
\ee
where $\kappa_n$ are the $n$-th cumulant of the probability distribution of the thermodynamic work. This cumulant series expansion agrees with the Second Law of Thermodynamics. Indeed
\be
\beta W_{\rm irr} = \sum_{n=2}^{\infty}\frac{[i (i\beta)]^n}{n!}\kappa_n= \ln{\phi(i\beta)}+\beta \langle W\rangle,
\ee
which leads to Jarzynski's equality~\cite{jarzynski1997}
\be
\langle e^{-\beta W}\rangle = e^{-\beta \Delta F}.
\ee
Observe that the expression~\eqref{eq:omegan} can be called a generalization of the fluctuation-dissipation relation. The idea is that by using such an expansion the diverging problem observed before for thermally isolated systems~\cite{naze2022series} will disappear.

\section{Nonlinear response of the irreversible work}

To generate the nonlinear response of $W_{\rm irr}$, the cumulants need to be expanded in the parameter perturbation of the process, and equal orders have to be summed up. Therefore
\be
\omega_n = \sum_{m=0}^\infty \omega_{nm}=\sum_{m=0}^\infty\frac{(-\beta)^n}{n!} \kappa_{nm},
\ee
where $\omega_{nm}$ and $\kappa_{nm}$ are the $m$-th term of the series expansion in the parameter perturbation of $\omega_n$ and $\kappa_n$. The $m$-th term of the series expansion of $W_{\rm irr}$ in the parameter perturbation is
\be
\beta W_{\rm irr}^{(m)} = \sum_{n=2}^{\infty}\omega_{nm}=\sum_{n=2}^{m}\frac{(-\beta)^n}{n!} \kappa_{nm},
\label{eq:wirr_m}
\ee
which can be called a generalization of the fluctuation-dissipation relation in the context of nonlinear response theory. To illustrate the result with a simple example, the linear response of $W_{\rm irr}$ is
\be
W_{\rm irr}^{(2)} = \frac{\beta}{2} \kappa_{22}=\frac{\beta}{2} \sigma^2_2,
\ee
where the term $\sigma^2_2$ is the variance of the work calculated until its second order in the parameter perturbation. This result has already been predicted~\cite{naze2023optimal}.

\section{Generalized relaxation functions}

The calculation of cumulants is not simple, and considering expansions in the parameter perturbation requires much more attention. In the next paragraphs, I will translate this procedure into the calculation of generalized relaxation functions, as is done in linear response theory~\cite{kubo2012statistical}.

To start, cumulants can be defined in terms of the $i$-th raw moments $\mu_i$ of the probability distribution
\be
\kappa_n=\sum_{i=1}^n(-1)^{i-1}(i-1)!\mathcal{B}_{n,i}(\mu_1,...,\mu_{n-i+1}),
\ee
where $\mathcal{B}_{n,i}$ are the partial Bell polynomials. In particular, the $m$-th expansion term $\kappa_{nm}$ will be expressed as
\be
\kappa_{nm}=\sum_{i=1}^n\sum_{m_1+...+m_{(n-i+1)}=m}(-1)^{i-1}(i-1)!\mathcal{B}_{n,i}\left(\mu_{1m_1},...,\mu_{(n-i+1)m_{n-1+1}}\right).
\label{eq:kappa_nm}
\ee
Now each raw moment has to be expanded in a series of the parameter perturbation
\be
{(-\beta)}^n\mu_n = {(-\beta)}^n\langle W^n\rangle,
\ee
which furnishes
\be
{(-\beta)}^n\mu_n = \int_0^\tau ... \int_0^\tau {(-\beta)}^n\left\langle \prod_{j=1}^n \mathcal{H}_1(t_j)\right\rangle \dot{g}(t_1)...\dot{g}(t_n)dt_1...dt_n.
\ee
The term on the average is calculated at different times. In this case, I consider that the probability distribution has the form
\be
\rho = \rho(z_{t_1},...,z_{t_n},t_1,...,t_n),
\ee
such that
\be
\int_{\Gamma_1}...\int_{\Gamma_n}\rho(z_{t_1},...,z_{t_n},t_1,...,t_n)dz_{t_1}...dz_{t_n}=1.
\ee
Demanding that the probability distribution should be invariant over time, we have
\be
d\rho = \sum_{i=1}^{n}\left(\frac{d \rho}{d t_i}\right)dt_i=0,
\ee
meaning
\be
\frac{d \rho}{d t_i} = 0 \quad \rightarrow \quad \frac{\partial \rho}{\partial t_i}=-\mathcal{L}_i\rho
\ee
The solution must satisfy each of Liouville's equations, requiring that the probability distribution starts from the canonical ensemble. Let us try the most trivial solution
\be
\rho(z_{t_1},...,z_{t_n},t_1,...,t_n)=\rho_1(z_{t_1},t_1)...\rho_n(z_{t_n},t_n),
\ee
with
\be
\rho(z_0,...,z_0,0,...,0)=\rho_c(z_0).
\ee
Under such circumstances, the moment will be
\be
{(-\beta)}^n\mu_n = \int_0^\tau ... \int_0^\tau {(-\beta)}^n\prod_{j=1}^n  \left\langle\mathcal{H}_1(t_j)\right\rangle_j \dot{g}(t_1)...\dot{g}(t_n)dt_1...dt_n,
\ee
where the index $j$ in the average indicates the $j$-th probability distribution of $\rho$. Using Liouville's theorem, one has
\be
{(-\beta)}^n\mu_n = \int_0^\tau ... \int_0^\tau {(-\beta)}^n\prod_{j=1}^n  \left\langle\mathcal{H}_1(0)\right\rangle_j \dot{g}(t_1)...\dot{g}(t_n)dt_1...dt_n.
\ee
Applying now the series expansion in the parameter perturbation
\be
{(-\beta)}^n\mu_n = \int_0^\tau ... \int_0^\tau {(-\beta)}^n\prod_{j=1}^n \sum_{i=0}^{\infty} \left\langle\mathcal{H}_1(0)\right\rangle_{ji} \dot{g}(t_1)...\dot{g}(t_n)dt_1...dt_n.
\ee
one can rewrite in the following form
\be
{(-\beta)}^n\mu_n = \sum_{i_1=0}^{\infty}...\sum_{i_n=0}^{\infty}\int_0^\tau ... \int_0^\tau {(-\beta)}^n\prod_{j=1}^n \left\langle\mathcal{H}_1(0)\right\rangle_{ji_j}\Bigg|_{i_1,...,i_n} \dot{g}(t_1)...\dot{g}(t_n)dt_1...dt_n.
\ee
The $m$-th term of the $n$-th moment will be
\be
{(-\beta)}^n\mu_{nm} = \sum_{n+i_1+...+i_n=m}\int_0^\tau ... \int_0^\tau {(-\beta)}^n\prod_{j=1}^n \left\langle\mathcal{H}_1(0)\right\rangle_{ji_j}\Bigg|_{i_1,...,i_n} \dot{g}(t_1)...\dot{g}(t_n)dt_1...dt_n.
\label{eq:mu_nm}
\ee
To each one of those terms can be put in terms of generalized relaxation functions. If all $i_j=0$, one has
\be
{(-\beta)}^n\prod_{j=1}^n\left\langle\mathcal{H}_1(0)\right\rangle_{j0}=\beta[\Psi_{n0}(t_1,...,t_{n})+\mathcal{C}_{n0}],
\ee
where
\be
-\Psi_{n0}(t_1,...,t_n)={(-\beta)}^{n-1}\left[\left\langle \prod_{j=1}^n e^{\mathcal{L}_0 t_j}\mathcal{H}_1(0)\right\rangle_0-\mathcal{C}_{n0}\right],
\ee
\be
\mathcal{C}_{n0}=\langle \mathcal{H}_1(0)\rangle_0^n.
\ee
If some $i_j\ge 1$, the calculation becomes more involved. Indeed
\be
{(-\beta)}^n\prod_{j=1}^n\left\langle\mathcal{H}_1(0)\right\rangle_{ji_j}\Bigg|_{i_1,...,i_n}=\left\langle(-\beta)^n\prod_{j=1}^n\mathcal{K}_{ji_j}(\Delta_{i_j},...,\Delta_1)\mathcal{H}_1(0)\right\rangle_0.
\label{eq:yes}
\ee
where the operator acts as
\be
\mathcal{K}_{ji_j}(\Delta_{i_j},...,\Delta_1)(\bullet)=(-1)^{i_j}\prod_{k=1}^{i_j}\int^{s_k}_0\mathcal{L}_1(0)e^{-\mathcal{L}_0\Delta_k}(\bullet)g(s_{k-1})ds_{k-1},
\ee
with $\Delta_k=s_k-s_{k-1}$ and $\Delta_{i_j}=t_j-s_{i_j-1}$. Observe that 
\be
\mathcal{L}_1(0)=\frac{d}{ds_{k-1}},
\ee
from which, performing an integration by parts, one has
\be
\mathcal{K}_{ji_j}(\Delta_{i_j},...,\Delta_1)(\bullet)=(-1)^{i_j}\prod_{k=1}^{i_j}\left[(\bullet)g(s_k)-\int^{s_k}_0 [e^{-\mathcal{L}_0\Delta_k}(\bullet)+\mathcal{C}_{jk}]\dot{g}(s_{k-1})ds_{k-1}\right],
\ee
or also
\be
\mathcal{K}_{ji_j}(\Delta_{i_j},...,\Delta_1)(\bullet)=\prod_{k=1}^{i_j}\left[\int^{s_k}_0 [(e^{-\mathcal{L}_0\Delta_k}-1)(\bullet)+\mathcal{C}_{jk}]\dot{g}(s_{k-1})ds_{k-1}\right].
\ee
Returning to Eq.~\eqref{eq:yes}
\be
{(-\beta)}^n\prod_{j=1}^n\left\langle\mathcal{H}_1(0)\right\rangle_{ji_j}\Bigg|_{i_1,...,i_n}=\left\langle(-\beta)^n\prod_{j=1}^n\prod_{k=1}^{i_j}\left[\int^{s_k}_0 [(e^{-\mathcal{L}_0\Delta_k}-1)\mathcal{H}_1(0)+\mathcal{C}_{jk}]\dot{g}(s_{k-1})ds_{k-1}\right]\right\rangle_0,
\ee
which is the same as
\be
{(-\beta)}^n\prod_{j=1}^n\left\langle\mathcal{H}_1(0)\right\rangle_{ji_j}\Bigg|_{i_1,...,i_n}=\left\langle(-\beta)^n\prod_{j=1}^n\int^{t_j}_0...\int^{s_1}_0\prod_{k=1}^{i_j}\left[ [(e^{-\mathcal{L}_0\Delta_k}-1)\mathcal{H}_1(0)+\mathcal{C}_{jk}]\right]\dot{g}(s_{i_j-1})...\dot{g}(s_0)ds_{i_j-1}...ds_0\right\rangle_0.
\ee
Observe that the property of switching the product and integrals can be performed one more time, leading to
\be
{(-\beta)}^n\prod_{j=1}^n\left\langle\mathcal{H}_1(0)\right\rangle_{ji_j}\Bigg|_{i_1,...,i_n}=\left(\prod_{j=1}^n\int^{t_j}_0...\int^{s_1}_0\right)\beta\Psi_{ni_1...i_n}(\{\Delta_1\}_j,...,\{\Delta_n\}_j)\left(\prod_{j=1}^n\dot{g}(s_{i_j-1})...\dot{g}(s_0)ds_{i_j-1}...ds_0\right),
\ee
where
\be
\beta\Psi_{ni_1...i_n}(\{\Delta_1\}_j,...,\{\Delta_n\}_j)=(-\beta)^{n}\left\langle \prod_{j=1}^n\prod_{k=1}^{i_j} [(e^{-\mathcal{L}_0\Delta_{k}}-1)\mathcal{H}_1(0)+\mathcal{C}_{jk}]\right\rangle_0,
\ee
The relaxation function $\Psi_{n\{i_1...i_n\}}(\{\Delta_1\}_j,...,\{\Delta_n\}_j)$ is the sum over the permutations of the set $\{i_1,...,i_n\}$ of $\Psi_{ni_1...i_n}(\{\Delta_1\},...,\{\Delta_n\})$
\be
\Psi_{n\{i_1...i_n\}}(\{\Delta_1\}_j,...,\{\Delta_n\}_j)=\sum_{n+i_1+...+i_n=m}\Psi_{ni_1...i_n}(\{\Delta_1\}_j,...,\{\Delta_n\}_j).
\ee
To determine the terms $\mathcal{C}_{jk}$, I use the criterion that the generalized relaxation functions should nullify when $\Delta_j\rightarrow \infty$. This will guarantee that at each expansion the moment will nullify at infinity, and also each cumulant and term of the irreversible work. In that limit, the canonical ensemble decorrelates at each random variable. Therefore
\be
\beta\Psi_{ni_1...i_n}(\{\infty\}_j,...,\{\infty\}_j)=(-\beta)^{n} \prod_{j=1}^n\prod_{k=1}^{i_j} \left\langle[(e^{-\mathcal{L}_0\Delta_{k}}-1)\mathcal{H}_1(0)+\mathcal{C}_{jk}]\right\rangle_0=0\quad \rightarrow \quad \mathcal{C}_{jk}=0.
\ee
Therefore, the relaxation function is the sum over the permutations of the set $\{i_1,...,i_n\}$ of the following function
\be
\beta\Psi_{ni_1...i_n}(\{\Delta_1\}_j,...,\{\Delta_n\}_j)=(-\beta)^{n}\left\langle \prod_{j=1}^n\prod_{k=1}^{i_j} [(e^{-\mathcal{L}_0\Delta_{k}}-1)\mathcal{H}_1(0)]\right\rangle_0.
\ee

\section{Examples}

The result seems to be consistent. Indeed, we recover linear response case, where $n=2$ and $i_1,i_2=0$,
\be
\Psi_{2\{00\}}(t_1,t_2)=\beta\left\langle\mathcal{H}_1(t_1)\mathcal{H}_1(t_2)\right\rangle_0-\beta\left\langle\mathcal{H}_1(0)\right\rangle_0^2.
\ee
Considering $n=3$ and $i_1,i_2,i_3=0$ one has
\be
\Psi_{3\{000\}}(t_1,t_2,t_3)=-\beta^2\left\langle\mathcal{H}_1(t_1)\mathcal{H}_1(t_2)\mathcal{H}_1(t_3)\right\rangle_0+\beta^2\left\langle\mathcal{H}_1(0)\right\rangle_0^3.
\ee
For its turn, when $n=2$ and $i_1,i_2=1$, one has
\be
\Psi_{2\{11\}}(t_1,t_2)=\beta\left\langle(\mathcal{H}_1(t_1)-\mathcal{H}_1(0))(\mathcal{H}_1(t_2)-\mathcal{H}_1(0))\right\rangle_0,
\ee
Also, when $n=3$ and $i_1,i_2,i_3=1$, one has
\be
\Psi_{3\{111\}}(t_1,t_2,t_3)=\beta^2\left\langle(\mathcal{H}_1(t_1)-\mathcal{H}_1(0))(\mathcal{H}_1(t_2)-\mathcal{H}_1(0))(\mathcal{H}_1(t_3)-\mathcal{H}_1(0))\right\rangle_0,
\ee
Observe now for $n=2$ with $i_1=0$ and $i_2=1$
\be
\Psi_{2\{01\}}(t_1,t_2)=\beta\left\langle\mathcal{H}_1(t_1)(\mathcal{H}_1(t_2)-\mathcal{H}_1(0))\right\rangle_0+\beta\left\langle\mathcal{H}_1(t_2)(\mathcal{H}_1(t_1)-\mathcal{H}_1(0))\right\rangle_0,
\ee
Finally, consider $n=2$ with $i_1=0$ and $i_2=2$. One has
\be
\Psi_{2\{01\}}(t_1,\{s_1,s_2\})=\beta\left\langle\mathcal{H}_1(t_1)(\mathcal{H}_1(s_1)-\mathcal{H}_1(0))(\mathcal{H}_1(s_2)-\mathcal{H}_1(0))\right\rangle_0+\beta\left\langle\mathcal{H}_1(s_1)(\mathcal{H}_1(t_1)-\mathcal{H}_1(0))(\mathcal{H}_1(t_2)-\mathcal{H}_1(0))\right\rangle_0,
\ee
Observe now the role of each index in the generalized relaxation function: the index $n$ indicates the number of terms that are multiplied in the average, while the index $i_j$ indicates the number of differences $(\mathcal{H}_1(t)-\mathcal{H}_1(0))$ involved in each one of those terms, except when $i_j=0$, where it becomes $\mathcal{H}_1(t)$. The sum over the permutations of $\{i_1,...,i_n\}$ indicates that the generalized relaxation functions are symmetric under the change of the vectors $\{\Delta_j\}$. Also, when there is $i_j\neq 0$, the constant equals zero. The sum $n+i_1+...+i_n$ indicates the order in the perturbation in the $n$-th moment. 

Observe that the calculation of the relaxation function is nothing more than a linear combination of the terms
\be
\mathcal{R}_{uv}=\left\langle\left(\prod_{i=1}^u\mathcal{H}_1(t_i)\right)\left(\prod_{i=1}^v\mathcal{H}_1(0)\right)\right\rangle_0,
\ee
with $u,v\ge 0$. These terms produce a relaxation matrix $\mathcal{R}(m)$ of order $m\times m$ for an approximation until the $m$-th order in the parameter perturbation. Because of properties from the average, the number of effective terms to be calculated are $m^2-m$.

\section{Procedure to calculate the approximation of the irreversible work}

The procedure to calculate the approximation of the irreversible work until its $n$-th order in the parameter perturbation is
\begin{itemize}
    \item Choose $n\ge 2$;
    \item Consider $m=2$;
    \item Identify the cumulants $\kappa_{l}$, with $2\le l\le m$;
    \item Identify the moments $\mu_p$, with $1\le p\le l$, that compose the cumulants $\kappa_l$;
    \item Calculate the remaining terms of relaxation matrix $\mathcal{R}(m)$;
    \item Calculate from $\mathcal{R}_{uv}$ all $\Psi_{p\{i_1,...,i_p\}}$, with $p+i_1+...+i_p=m$;
    \item Use Eq.~\eqref{eq:mu_nm} to calculate $\mu_{pm}$;
    \item Use Eq.~\eqref{eq:kappa_nm} to calculate $\kappa_{lm}$;
    \item Use Eq.~\eqref{eq:wirr_m} to calculate to $m$-th order of the irreversible work;
    \item If $m=n$ is true, go to the next step, otherwise repeat the process from the third step considering $m+1$;
    \item Sum up all the $n-1$ terms calculated before to find the approximation until $n$-th order.
\end{itemize}

\section{Final remarks}

On this work, I developed a nonlinear response theory for the irreversible work. The use of cumulants series expansion is an attempt to avoid diverging behavior in this case, since it establishes generalized relaxation functions that nullify at infinity by the choice of constants that provide such behavior. Providing however an example of a system performing an isothermal process until higher orders to verify such behavior is a herculeous work. A new version of this work will be shared when such an example is calculated. 

\bibliography{NRCS.bib}
\bibliographystyle{apsrev4-2}

\end{document}